\documentclass[conference]{IEEEtran}

\ifCLASSINFOpdf

\else

\fi

\usepackage{amsfonts}
\usepackage{graphicx}
\usepackage{url}
\usepackage{multirow}
\usepackage{stmaryrd}
\usepackage{amsmath }
\usepackage{multirow}
\usepackage{xcolor}
\usepackage{lipsum}

\newcommand{\maybe}[1]{{\color{red}}}
\newcommand{\repourl}{https://github.com/AICPS/PLCBEAD\_PLCEmbed}

\begin{document}
\pagestyle{plain} 
\title{Bridging the PLC Binary Analysis Gap: A Cross-Compiler Dataset and Neural Framework for Industrial Control Systems}

\author{
\IEEEauthorblockN{%
  Yonatan~Gizachew~Achamyeleh\IEEEauthorrefmark{1},
  Shih-Yuan~Yu\IEEEauthorrefmark{1},
  Gustavo~Quir\'os~Araya\IEEEauthorrefmark{2},
  Mohammad~Abdullah~Al~Faruque\IEEEauthorrefmark{1}
}
\IEEEauthorblockA{\IEEEauthorrefmark{1}University of California, Irvine, Irvine, CA, USA \\
\texttt{\{yachamye,shihyuay,alfaruqu\}@uci.edu}}
\IEEEauthorblockA{\IEEEauthorrefmark{2}Siemens Technology, Princeton, NJ, USA \\
\texttt{gustavo.quiros@siemens.com}}
}


\maketitle

\begin{abstract}
Industrial Control Systems (ICS) rely heavily on Programmable Logic Controllers (PLCs) to manage critical infrastructure, yet analyzing PLC executables remains challenging due to diverse proprietary compilers and limited access to source code.
To bridge this gap,
we introduce PLC-BEAD, a comprehensive dataset containing 2431 compiled binaries from 700+ PLC programs across four major industrial compilers (CoDeSys, GEB, OpenPLC-V2, OpenPLC-V3).
This novel dataset uniquely pairs each binary with its original Structured Text source code and standardized functionality labels, enabling both binary-level and source-level analysis.
We demonstrate the dataset's utility through PLCEmbed, a transformer-based framework for binary code analysis that achieves 93\% accuracy in compiler provenance identification and 42\% accuracy in fine-grained functionality classification across 22 industrial control categories.
Through comprehensive ablation studies, we analyze how compiler optimization levels, code patterns, and class distributions influence model performance.
We provide detailed documentation of the dataset creation process, labeling taxonomy, and benchmark protocols to ensure reproducibility.
Both PLC-BEAD and PLCEmbed are released as open-source resources to foster research in PLC security, reverse engineering, and ICS forensics, establishing new baselines for data-driven approaches to industrial cybersecurity.

\end{abstract}

%
\IEEEpeerreviewmaketitle

\section{Introduction}
Industrial Control Systems (ICS) rely heavily on \textit{Programmable Logic Controllers (PLCs)} to manage critical infrastructure such as manufacturing, power generation, and transportation~\cite{erickson2016programmable, alphonsus2016review}.
Despite the advent of newer systems, many industrial sites continue to operate legacy PLCs that lack up-to-date documentation and source code~\cite{wang2023survey}. This creates significant challenges for security analysis and maintenance, particularly in facilities that must remain operational around the clock~\cite{rupprecht2021concepts,keliris2019icsref, chivilikhin2020automatic}.
High-profile incidents like Stuxnet and Triton demonstrate how attackers can target the PLC layer to disrupt physical processes with severe real-world consequences~\cite{giles2019triton,makrakis2021vulnerabilities}.
In these cases, threat actors exploited vulnerabilities in the toolchain or the deployed PLC program.
Such attacks underscore the urgent need for methods to inspect and analyze PLC executables even when source code is unavailable~\cite{giles2019triton,makrakis2021vulnerabilities,keliris2019icsref, wang2023survey}.

Binary code analysis offers a promising solution by examining executables directly without requiring source code access~\cite{song2008bitblaze, shoshitaishvili2016sok, 7985685, yu2023cfg2vec}. 
This technique examines the executable itself to uncover structure, behavior, and potential vulnerabilities.
However, while binary analysis is well-established for standard platforms like Linux or Windows, its application to PLCs remains challenging.
PLCs use vendor-specific compilers~\cite{codesys, openplc, geb, step7} that produce proprietary binary formats, making traditional analysis tools ineffective~\cite{keliris2019icsref, chang2018constructing, 7006408}. 
The challenge is compounded by limited access to source code and documentation in legacy systems, along with diverse programming languages and standards like IEC 61131-3~\cite{international3international}. Most critically, the field lacks standardized datasets for developing machine learning-based analysis techniques.

To bridge this gap, 
we introduce \textbf{PLC-BEAD (PLC Binary Evaluation and Analysis Dataset)}, a comprehensive dataset containing \textit{2431 compiled binaries} from \textit{over 700 PLC programs} across four major industrial compilers. 
This novel dataset uniquely pairs each binary with its original Structured Text source code and standardized functionality labels, \textit{enabling} both \textit{binary-level} and \textit{source-level} analysis for machine learning research. 
The dataset covers binaries compiled using CoDeSys~\cite{codesys}, GEB~\cite{geb}, OpenPLC-V2~\cite{openplc}, and OpenPLC-V3~\cite{openplc}, with programs derived from the industry-standard OSCAT library~\cite{oscat}. 
Each binary is labeled across 22 functionality categories, providing rich metadata for supervised learning tasks.

To demonstrate the dataset's utility, we present \textbf{PLCEmbed}, a transformer-based embedding framework that ingests raw bytes of PLC binaries to accomplish tasks such as \textbf{toolchain provenance identification} (i.e., discovering which compiler produced a given binary) and \textbf{functionality classification} (e.g., distinguishing a timer routine from a network communication block). These tasks are vital in \textbf{ICS digital forensics}, where investigators often must quickly determine whether a suspicious binary matches known legitimate code, or whether it stems from a vulnerable or maliciously modified compiler.

\textbf{Contributions.} Our work provides: \begin{itemize} \item \textbf{PLC-BEAD Dataset.} The first well-documented, open-source PLC binary dataset, comprising over 700 programs compiled with four different toolchains.

    \item \textbf{PLCEmbed.} A framework that leverages machine learning for multi-class classification of binary identity. 

    \item \textbf{Benchmarks \& Findings.} Experimental results show PLCEmbed can extract forensic information from various vendor-specific PLC binaries. PLCEmbed achieves 93\% accuracy for toolchain provenance and about 42\% for functionality classification.
    
    \item \textbf{Open-Source Release \& Future Impact.} Both the dataset and the PLCEmbed code are publicly released. Our repository is available at \url{\repourl}.

\end{itemize}

Overall, our work bridges an important gap in ICS security research by providing an all-in-one PLC binary dataset and a flexible embedding approach that can adapt to new compiler versions or vendor libraries.
Our findings highlight the inherent complexity of PLC binary analysis, while showing how data-driven methods, guided by open benchmarks, can strengthen both academic research and industrial cybersecurity practices.


\section{Related Works and Background}
\label{sec:related}
Binary analysis has long been essential in software security, enabling vulnerability assessment and malware detection without source code access. For general-purpose software, established tools~\cite{hex2017ida, radare2book, nsfghidra} and techniques leverage standardized formats like ELF on Linux or PE on Windows. However, PLC binary analysis presents unique challenges due to proprietary formats and diverse industrial requirements.

\subsection{PLC Binary Analysis Challenges}
PLCs serve as a cornerstone in various industrial sectors due to their extensive integration and controlling capability in complex automation systems. Engineers develop PLC programs using standardized languages defined by the International Electrotechnical Commission (IEC 61131-3)~\cite{international3international}, with Structured Text (ST) being a high-level programming language syntactically resembling Pascal. These programs are then compiled using vendor-specific toolchains into proprietary binary formats.

In practice, \textbf{PLC toolchains} vary widely. Each vendor typically provides its own proprietary compiler and Integrated Development Environment (IDE) for programming. 
Examples include CoDeSys, Siemens STEP7, GEB, and OpenPLC~\cite{codesys,step7,geb,openplc}. Although these vendor-specific environments simplify the engineering workflow, they also produce binary code in nonstandard formats. 
This heterogeneity complicates efforts to develop universal analysis or reverse-engineering tools.

\subsection{Machine Learning for Binary Analysis}
Recent advances in machine learning (ML) have shown promise in binary analysis tasks such as functionality classification and compiler identification~\cite{pizzolotto2020identifying, zuo2018neural,marcelli2022machine, ding2019asm2vec, duan2020deepbindiff,yu2023cfg2vec,  gao2018vulseeker, marcelli2022machine}. Neural architectures like Convolutional Neural Networks (CNNs) and Transformers have demonstrated effectiveness in learning patterns from raw binary data. However, these approaches typically rely on large, labeled datasets, a resource notably absent in the PLC domain.

Prior, non-ML, PLC binary analysis efforts have focused on specific toolchains, particularly CoDeSys~\cite{keliris2019icsref, ICSFuzz}. 
For example, ICSREF provides a framework for automated reverse engineering of CoDeSys-generated binaries, while ICSFuzz enables fuzzing of control applications~\cite{keliris2019icsref}. 
However, these solutions are limited to single compilers or narrow subsets of PLC platforms. The lack of comprehensive, cross-compiler datasets has hindered the development of more general machine-learning approaches.

\subsection{Need for Standardized PLC Datasets}
The scarcity of publicly available, well-documented PLC binary datasets presents a significant barrier to research. Several factors contribute to this gap: First, PLC systems often control critical infrastructure, making organizations reluctant to share operational code. Second, intellectual property concerns restrict 
the free exchange and availability of these resources and stifling the potential for broader, more inclusive research and understanding of PLC systems.
Third, the diversity of vendor-specific formats requires expertise across multiple industrial platforms to create representative datasets. The absence of data-sharing incentives further exacerbates the challenge.
These challenges underscore the need for an open, comprehensive dataset that enables:

(1) Development of machine learning models for cross-compiler binary analysis

(2) Standardized evaluation of PLC binary analysis techniques

(3) Research into industrial control system security and forensics

The following sections introduce PLC-BEAD, addressing these needs through a carefully curated collection of PLC binaries spanning multiple compilers and industrial applications.

\section{\textit{PLC-BEAD} Dataset}
\label{sec:plcbead}

\subsection{Dataset Overview and Design Goals}
\label{subsec:overview}
\textbf{PLC-BEAD} is a comprehensive dataset encompassing over 700 unique PLC programs and 2,431 binaries, meticulously compiled using four different compilers. 
Each program in the dataset is accompanied by its source code written in Structured Text (ST) programming language, following the IEC 61131-3 standard~\cite{international3international}. These programs range from basic control mechanisms like timers and flip-flops to sophisticated algorithms for data processing and network communication.
The dataset is designed to capture the diversity of PLC program development in industrial settings, with binaries generated using four distinct compilers:

\begin{itemize}
    \item \textbf{CoDeSys} \cite{codesys}: A popular platform that many industrial vendors adopt~\cite{keliris2019icsref}. It produces a distinctive binary layout with dedicated sections for variable initialization and runtime system data.
    \item \textbf{GEB} \cite{geb}: A closed-source compiler often used in certain industrial sectors for building robust automation logic.
    \item \textbf{OpenPLC-V2} \cite{openplc}: An open-source solution that follows the IEC 61131-3 standard and is valued for its simplicity.
    \item \textbf{OpenPLC-V3} \cite{openplc}: A more modern iteration of OpenPLC with additional optimizations. 
\end{itemize}

These represent a mix of proprietary and open-source solutions that are widely used in industry and research~\cite{keliris2019icsref,7006408}.
Including these four compilers helps illustrate how the same ST source can produce substantially different binary layouts, enabling the study of unique compiler-specific patterns in binaries and facilitating the development of universal binary analysis tools.
To our knowledge, \textit{no other open dataset} offers such a combination of breadth (multi-compiler coverage) and depth (source-binary pairs, extensive functionalities) for PLCs.

\paragraph{Why an Integrated Source-Binary Dataset.}
Providing both source and binary artifacts in PLC-BEAD enables two complementary research directions. Researchers aiming to develop \textit{binary-focused} approaches (e.g., for ICS digital forensics or vulnerability scans) can rely solely on the compiled executables to evaluate how well automated methods perform without high-level code. Conversely, others may want to align their binary analyses with known ground-truth logic at the source level in order to map discovered vulnerabilities back to specific function blocks or instructions. This integrated approach offers flexibility in exploring questions such as:
\begin{itemize}
\item How do different compilers implement the same control algorithm, and what patterns or “fingerprints” are visible in the final binaries? 
\item Can a trained model reliably infer functionality (e.g., “timer logic” vs. “PID control loop”) solely from raw bytes? 
\item Does the same snippet of ST code produce semantically similar but structurally different binaries under different optimization settings or hardware targets? \end{itemize}

\paragraph{Potential Impact.}
Openly available datasets in computer vision and natural language processing have accelerated progress by enabling reproducibility and fostering collaborative benchmarking.
We believe a similar approach can empower ICS binary analysis research, where industrial secrecy and regulatory requirements often constrain reproducibility.
Our hope is that PLC-BEAD will:
\begin{itemize}
    \item Illuminate compiler-specific differences that may introduce security weaknesses or vulnerabilities in PLC binaries.
    \item Provide a standard platform for comparing machine learning models aimed at tasks such as toolchain provenance identification, functionality classification, or anomaly detection in PLC firmware.
    \item Spur follow-up work in ICS digital forensics, where effective classification of suspicious PLC binaries and rapid understanding of their purpose could significantly reduce response times after an incident.
\end{itemize}

\begin{figure}[!ht]
    \centering
    \includegraphics[width=0.88\linewidth]{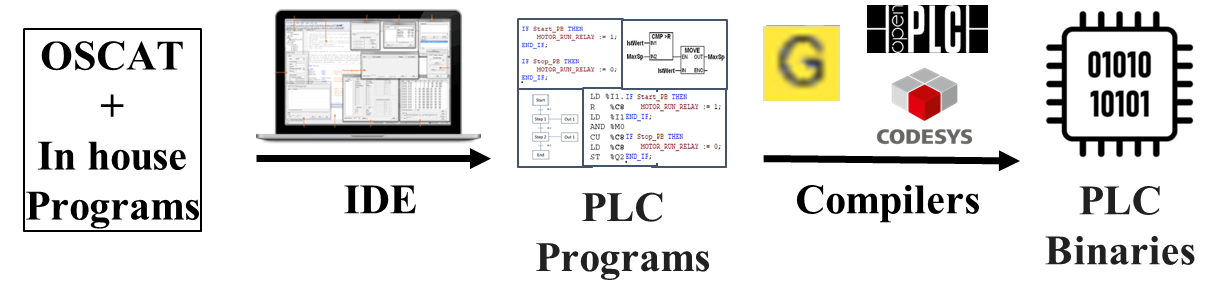}
    \caption{PLC-BEAD Dataset Construction Pipeline}
    \vspace{-8pt}
    \label{fig:compile}
\end{figure}

\subsection{Dataset Construction}
\label{subsec:construction}

The primary data source for PLC-BEAD is the \textit{OSCAT (Open Source Community for Automation Technology)} library~\cite{oscat}, a significant resource in PLC programming and automation~\cite{langmann2016aplc,
canedo2021arducode}. The library includes function blocks that range from arithmetic and signal processing to advanced building automation routines. 
OSCAT covers a broad spectrum of functions, from basic arithmetic to building automation modules, making it a practical starting point for reflecting real-world industrial requirements. 
Our aim was to capture a broad set of use cases that an engineer might encounter in industrial environments, while still keeping the dataset manageable.

\paragraph{Compilation Procedure.} Fig.~\ref{fig:compile} depicts the compilation process. After developing the PLC programs, we proceeded to compile them using the selected toolchains to obtain their corresponding binary files. However, due to variations in built-in libraries and primitive functions, some programs did not compile across all four compilers. 
For example, a string-handling snippet called \texttt{CAPITALIZE.ST} could be compiled in \textit{CoDeSys} but failed under \textit{OpenPLC-V3}. This is because the \textit{OpenPLC-V3} compiler does not support the primitive function, \texttt{\text{GET}\_\text{CHAR}()}. 
Similar compiler-specific constraints emerged across different functionality domains. 
In cases where only minor tweaks were required (for instance, replacing a function call with an equivalent), we performed these adjustments. 
For example, given that the language parsers for CoDeSys and GEB differ substantially from those used by OpenPLC, modifications to the ST programs were necessary to prevent syntax errors during the compilation process.
If a program demanded extensive rewrites to work on a particular compiler, we excluded it from that compiler’s compilation list to avoid changing its core logic. 


\begin{table}[!t]
    \centering \renewcommand{\arraystretch}{1.5}
    \caption{The data statistics of the \textit{PLC-BEAD} dataset.}
    \begin{tabular}{|l|l|l|l|l|}
    \hline
        \multirow{2}{*}{\# of Programs} & \multicolumn{4}{|c|}{\# of Binaries } \\ \cline{2-5}
        ~ & \textit{CoDeSys} & \textit{GEB} & \textit{OpenPLC-V2} & \textit{OpenPLC-V3} \\ \hline
        729 & 555 & 617 & 619 & 640 \\ \hline
    \end{tabular}
\vspace{-10pt}
    \label{tab:datastat}
\end{table}
\vspace{3pt}

\paragraph{Compilation Result.} Table~\ref{tab:datastat} summarizes how many of the 729 ST programs were successfully compiled by each toolchain. 
Specifically, \textit{OpenPLC-V2} compiled 619 programs, and \textit{OpenPLC-V3} compiled 640, while \textit{CoDeSys} and \textit{GEB} successfully handled 555 and 617 programs, respectively. 
These differences stem from language parsers and built-in library support that vary among the toolchains.
By the end of this process, we accumulated \textbf{2431 valid binaries} overall.

\subsubsection{Functionality Labeling}
\label{subsubsec:labeling}
One of the central goals of PLC-BEAD is to facilitate functionality-based analysis. 
For instance, security analysts might want to detect whether a suspicious binary is a timer routine or an actuator controller. 
Similarly, machine learning researchers might try to classify binaries based on high-level tasks. 
We therefore assigned each ST snippet to one of \textbf{22 functional categories}, derived from the OSCAT documentation~\cite{oscat} and common industrial practice (for example, “Timing,” “Math,” “Network,” “Building Control,” and others).

\paragraph{Annotation Procedure.}
The labeling was done manually by parsing the program’s file name and function block references. 
Some programs implemented multiple functions and did not fit neatly into a single label. 
In those cases, we used the most dominant function block or subroutine. 
For example, a snippet that primarily handled string manipulation but also included a small timer was labeled “String processing.” 
At least two team members confirmed all labels, who cross-checked the ST source to ensure consistency.

\begin{figure}[!t]
    \centering
    \includegraphics[width=0.9999\linewidth]{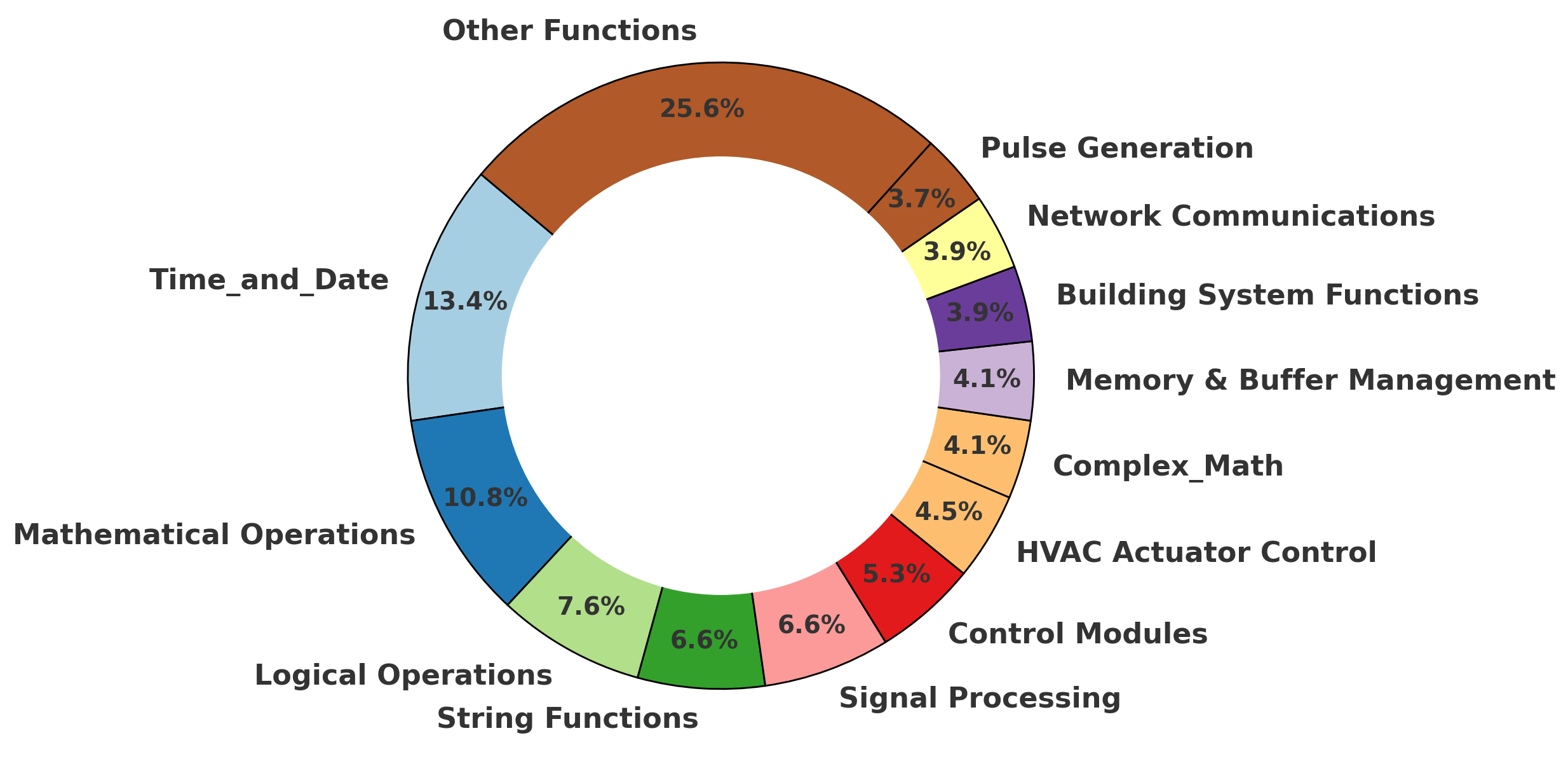}
    \caption{Distribution Of Functional Labels (Top 12 Categories + Others)}
    \vspace{-8pt}
    \label{fig:func_labels}
\end{figure}

Fig.~\ref{fig:func_labels} illustrates the distribution of these functionality labels, with “\texttt{Time\_and\_Date}” and “\texttt{Mathematical Operations}” blocks being among the most common. 
We note that some functional categories have fewer samples, indicating mild class imbalance that researchers should account for during training or evaluation.


\subsection{Dataset Organization}
\label{subsec:organization}

To facilitate navigation and reuse, the PLC-BEAD dataset is organized into a hierarchy that separates source files, compiled binaries, and supporting metadata. The top-level directory contains:

\begin{itemize}
    \item \textit{Source/}: All ST programs. Files are grouped by compiler compatibility. For instance, \texttt{Source/codesys/} includes ST files that successfully compile with CoDeSys.
    \item \textit{Binary/}:  The resulting binaries, arranged similarly by compiler.
    \item \textit{Metadata/}: CSV files describing success or failure status for each ST program file. This directory also contains the master index file mapping each ST code to its compiled binaries and assigned functionality labels.
    \item \textit{README.md}: Documentation to ensure it is accessible to researchers. This offers in-depth insights into the dataset's structure, clarifies the significance of each metadata field, and provides guidelines for leveraging the dataset effectively in various research contexts.
\end{itemize}

This organization ensures that researchers can easily navigate across multiple compilers and functionality categories, with clear visibility into which files and binaries belong to each class. We also have a data card for the dataset attached as supplementary material.


\subsection{Significance of Dataset Diversity}
Understanding binary data and its intricacies is fundamental in the context of PLC systems. 
Dataset diversity serves multiple critical purposes. First, it enhances static binary analysis, which focuses on examining binary code structure, control flow, and data flow. PLC-BEAD ensures analytical tools can interpret different vendor-specific binary formats and programming methodologies. Furthermore, diverse PLC binaries allow researchers to evaluate the limitations and accuracy of existing static analysis tools. By benchmarking these tools against a diverse dataset, researchers can identify areas for improvement and develop enhanced techniques that address the challenges presented by different PLC binaries.

The dataset's diversity is particularly crucial for ML applications. While ML techniques have shown promise in classifying and analyzing general-purpose software binaries~\cite{kim2018multimodal,shabtai2010automated,su2016deep,yu2023cfg2vec,xu2017spain}, the proprietary nature of PLC binaries, coupled with the lack of large, labeled datasets, has limited ML applications in the PLC domain. A diverse dataset that includes PLC binaries from multiple compilers facilitates training ML models that can generalize well across different PLC systems.

PLC-BEAD's structure and diversity aim to stimulate and support further research in PLC binary analysis. Its organization, metadata, and source code offer a robust platform for exploring PLC binaries, refining analysis techniques, and developing strong ML models for PLC system security and reliability. The dataset's comprehensive nature makes it particularly valuable for developing and evaluating cross-compiler binary analysis techniques, which are essential for handling the heterogeneous nature of industrial control systems.
\section{Binary Embedding Framework}
\label{sec:PLCEmbed}
This section introduces our proposed framework, \textbf{PLCEmbed}, a raw-byte embedding method for PLC binaries. 
PLCEmbed is designed to be compiler-agnostic and to handle heterogeneous binary formats without relying on specialized reverse-engineering tools.
We begin by outlining the overall approach and objectives (Sec.~\ref{subsec:plcembed_approach}), then formalize the two primary classification tasks that PLCEmbed addresses (Sec.~\ref{subsec:plcembed_tasks}).

\subsection{Overall Approach and Goals}
\label{subsec:plcembed_approach}

\paragraph{Context and Motivation.}
Most industrial control system (ICS) security and forensics efforts have historically focused on high-level source code analysis or network-layer monitoring. However, as discussed in Sec.~\ref{sec:related} and~\ref{sec:plcbead}, many legacy environments lack accessible source code and rely on diverse, proprietary compiler toolchains to produce PLC executables. 
Traditional reverse-engineering methods assume standard file formats, symbol tables, or specific CPU architectures, yet PLC binaries often break these assumptions\cite{song2008bitblaze,shoshitaishvili2016sok,keliris2019icsref, ICSFuzz}. 
Moreover, multiple compilers each introduce unique design choices, making it difficult to devise a one-size-fits-all reverse-engineering technique.
In response to these challenges, we developed PLCEmbed to operate \textit{directly on the raw bytes} of a PLC binary, minimizing the assumptions about its internal format.

\paragraph{High-Level Idea.}
Rather than parsing executable sections or relying on symbol information, PLCEmbed consumes the entire binary as a variable-length sequence of bytes.
A learnable embedding layer first transforms each byte into a vector representation, allowing the model to discover relationships between byte values automatically.
Subsequent neural network components (detailed in Sec.~\ref{subsec:plcembed_architecture}) capture both local patterns, such as repeated compiler inserts, and global dependencies that span distant parts of the binary.
By avoiding compiler-specific heuristics or file-structure assumptions, PLCEmbed can unify analysis across multiple toolchains.

\paragraph{ICS Security Relevance.}
Two central questions emerge when investigating suspicious binaries in Industrial Control Systems (ICS):

(1) \textit{Which toolchain produced this executable?}

(2) \textit{What functionality does it implement?}

Identifying the toolchain (for example, CoDeSys vs. GEB) can be crucial if a known vulnerability or exploit path exists for a particular compiler version.
Determining the functionality can verify whether the binary’s behavior aligns with the claimed control task (for example, a timer routine, network protocol handler, or building automation function).
PLCEmbed targets these questions by mapping raw bytes to higher-level insights, paving the way for automated forensics in ICS environments.

\paragraph{Goals of PLCEmbed.}
Building on the multi-toolchain dataset introduced in Sec.~\ref{sec:plcbead}, we aim to:
\begin{itemize}
    \item \textit{Provide a unified method} that handles binaries from different compilers without custom reverse-engineering routines.
    \item \textit{Facilitate classification tasks} such as toolchain provenance and functionality detection, which are vital to ICS forensics.
    \item \textit{Enable future ICS security research} by offering a reference architecture that is purely data-driven and thus extensible to emerging proprietary formats.
\end{itemize}

As we envision the development of more sophisticated tools in the future, with our dataset playing a pivotal role in this progression, our immediate goal is to establish an ML-based compiler-agnostic framework for binary analysis. 
In the following subsections, we formalize the classification tasks (Sec.~\ref{subsec:plcembed_tasks}), then present the detailed model design (Sec.~\ref{subsec:plcembed_architecture}).

\subsection{Problem Formulation and Classification Tasks}
\label{subsec:plcembed_tasks}
In order to clarify the PLCEmbed approach, we define the input space (raw binary bytes) and two separate classification objectives: \textbf{toolchain provenance} and \textbf{functionality}.

\paragraph{Input Representation.}
Let $X$ denote the set of all PLC binaries within our dataset.
Each binary $b \in X$ is represented as a sequence of bytes $\mathbf{x} = (x_1, x_2, \dots, x_L)$, where $x_i \in \{0,1,\dots,255\}$ and $L$ may vary across files.
When $L$ surpasses a maximum length (for example, 65,536 bytes), we truncate; when $L$ is shorter, we pad with a special token to indicate empty bytes.
Hence, every binary is mapped to a fixed-length sequence $\mathbf{x}$, which will be the input to our model.

\paragraph{Toolchain Provenance.}
We denote the toolchain label as $y_{\mathrm{tc}} \in \{\text{CoDeSys}, \text{GEB}, \text{OpenPLC-V2}, \text{OpenPLC-V3}\}$.
This classification objective seeks to identify which compiler produced the binary.
Formally, we learn a function $f_{\mathrm{tc}}: \mathbf{x} \mapsto y_{\mathrm{tc}}$ that assigns one of these four classes.
Identifying compiler origin is critical for ICS security because certain vulnerabilities might be exclusive to specific toolchains, and forensic analysts can narrow their search if they know the toolchain used \cite{keliris2019icsref}. 
Table~\ref{tab:vulstat} shows the numbers of vulnerabilities identified in well-known PLC development systems reported to the \textit{National Vulnerability Database} (NVD)~\cite{NVD} and the average \textit{Common Vulnerability Scoring System} (CVSS) severity of them.
When analyzing PLC binaries, knowledge of the toolchain provenance allows investigators to examine the origins of threats more effectively, especially if the toolchain in question has been previously associated with similar security issues. Furthermore, utilizing compiler-related information makes it possible to model the compilation chain, which can enhance defenses against cyberattacks by providing detailed semantic behavior insights ~\cite{ben2018detection,rahimian2015bincomp,otsubo2020o-glassesx,benoit2021binary, yang2022ratscope}.

\begin{table}[!hb]
    \centering
    \caption{The reported number of PLC toolchain vulnerabilities in NVD and the average of CVSS severity.}
    \begin{tabular}{c|c|c}
        \hline
        PLC Toolchain & \# of Vulnerabilities & Avg. CVSS Severity\\
        \hline
        \textit{CoDeSys} & 11 & 7.75 (High)\\
        OpenPLC & 2 & 7.10 (High)\\
        Siemens TIA Portal & 57 & 6.52 (Medium)\\
        TwinCAT & 6 & 7.52 (High)\\
        RSLogix & 10 & 6.84 (Medium)\\
        \hline
    \end{tabular}
    \label{tab:vulstat}
\end{table}

\begin{figure*}[!t]
    \centering
    \includegraphics[width=0.99\textwidth]{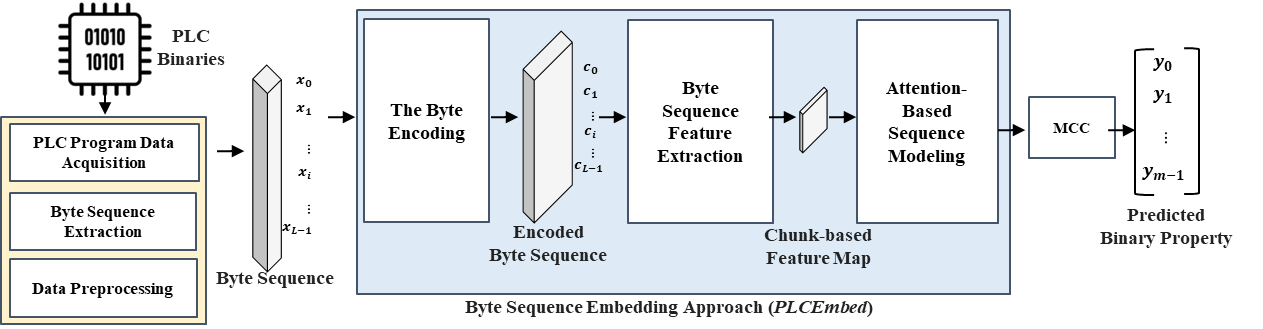} 

    \caption{The overall architecture of \textit{PLCEmbed}}
    \label{fig:archi}
\end{figure*}

\paragraph{Functionality Classification.}
A second classification task focuses on the functional category $y_{\mathrm{func}} \in \{\text{Timing}, \text{Network}, \dots\}$, comprising 22 labels derived from Sec.~\ref{subsubsec:labeling} of our dataset construction.
We learn a function $f_{\mathrm{func}}: \mathbf{x} \mapsto y_{\mathrm{func}}$ that predicts which of the 22 categories best describes the binary’s core behavior.
This capability assists ICS operators in verifying that a given binary indeed implements (for example) a building automation block rather than an unauthorized network-scanning routine or a maliciously modified PID controller.
This approach complements any available metadata from engineering workstations, audit logs, or network captures, allowing forensic teams to focus on files that deviate from expected patterns. It can also reduce manual hex inspections or partial reverse-engineering tasks by surfacing immediate leads in the early stage of an investigation.

\subsection{Model Architecture}
\label{subsec:plcembed_architecture}

Building on the problem formulation, we now present the detailed architecture of \textbf{PLCEmbed}. Fig.~\ref{fig:archi} offers a conceptual view of the pipeline, while the subsections below discuss each component. 

\subsubsection{Byte Encoding Layer}
\label{subsubsec:byte_embedding}
Each PLC binary is read as a sequence of bytes $\mathbf{x} = (x_1, x_2, \dots, x_L)$, where $x_i \in \{0,\dots,255\}$ and $L$ is fixed by truncation or padding.
We map each byte $x_i$ to a learnable embedding vector $\mathbf{e}_i \in \mathbb{R}^{d}$, where $d$ is a hyperparameter (for example, $d=256$).
We learn these embeddings jointly with the rest of the model, allowing the network to identify meaningful relationships among different byte values during training.
The resulting sequence $(\mathbf{e}_1, \mathbf{e}_2, \dots, \mathbf{e}_L)$ serves as the input to the subsequent layers.

\subsubsection{CNN-Based Local Feature Extraction}
\label{subsubsec:cnn_extractor}

After we obtain the byte embeddings, 
we use a one-dimensional convolutional layer to capture local patterns in the byte sequence~\cite{raff2018malware}. For example, many compilers insert characteristic blocks or repeated sequences at regular intervals. A CNN can exploit these short-range dependencies to generate a set of feature maps:
\begin{equation}
    \mathbf{h} = \text{Conv1D}(\mathbf{e}_1, \mathbf{e}_2, \dots, \mathbf{e}_L),
\end{equation}
where the kernel size and stride can be tuned to reflect typical repeated segments in PLC binaries. This extracted representation $\mathbf{h}$ emphasizes local byte patterns that may reveal compiler-specific “fingerprints.”

\subsubsection{Transformer for Global Context}
\label{subsubsec:transformer}
To capture global, long-range relationships within the byte sequence, we feed $\mathbf{h}$ into a transformer encoder~\cite{vaswani2017attention}. The encoder’s multi-head self-attention learns how different positions in the byte sequence relate to one another, allowing the model to link distant parts of the binary that might encode function calls or configuration blocks. While the CNN captures local cues, the transformer integrates these cues into a broader context vector.

\subsubsection{Classification Head}
\label{subsubsec:classification_head}

The final transformer output is a sequence of context vectors $(\mathbf{z}_1, \dots, \mathbf{z}_M)$, where $M \leq L$ depends on any pooling or downsampling used in the CNN stage. We apply a global pooling operation as the summary vector $\mathbf{z}_{\mathrm{summary}}$. A fully connected layer then maps $\mathbf{z}_{\mathrm{summary}}$ to class logits:
\begin{equation}
    \mathbf{p} = \mathrm{Softmax}\bigl(\mathbf{W}\,\mathbf{z}_{\mathrm{summary}} + \mathbf{b}\bigr).
\end{equation}
For \textit{toolchain provenance}, the size of $\mathbf{p}$ is 4, while for \textit{functionality classification}, it is 22 (matching the number of categories).

\subsubsection{CNN-Only Baseline}
\label{subsubsec:cnn_baseline}

Although the transformer encoder should help capture long-range dependencies, we also implement a simpler \textbf{CNN-only} baseline that omits the transformer layers. In this variant, the model processes byte embeddings through one or more convolutional blocks, then applies a final pooling and softmax layer. The CNN-only approach is expected to be more efficient but may struggle with binaries whose distinctive features are distributed across large code segments. 
This design is similar to prior raw-binary classifiers for general-purpose Linux binaries (e.g., MalConv~\cite{raff2018malware}), which detect local n-gram patterns but cannot directly capture long-range dependencies within a file.

\begin{figure}[!ht]
    \centering
    \includegraphics[width=1.0\linewidth]{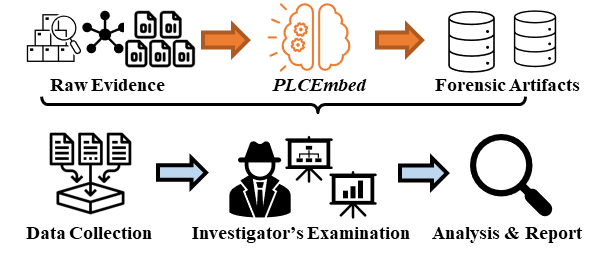}
    \caption{PLC DF process comprises four steps: data collection, examination, analysis, and reporting. To analyze security incidents, the investigator looks at raw evidence and leverages the forensic artifacts for the following stages. \textit{PLCEmbed} serves as a tool to help investigators extract information from raw evidence, making the process more efficient. }
    \label{fig:DF}
\end{figure}

\subsection{Application to PLC Digital Forensics}
\label{subsec:plc-forensics}

Fig.~\ref{fig:DF} illustrates the complete PLC digital forensics (DF) process, which comprises four steps: data collection, examination, analysis, and reporting~\cite{kent2006guide, karabiyik2018forensic}. In a typical investigation, the analyst first gathers raw evidence from PLC systems. During the examination stage, forensic artifacts are extracted from this raw data. Here, \textit{PLCEmbed} serves as a tool to automatically extract meaningful information from the raw bytes, making the subsequent analysis phase more efficient. By predicting both toolchain provenance and functionality, PLCEmbed assists investigators in verifying whether the behavior of an executable aligns with its claimed purpose. This capability streamlines the reporting process and helps focus further analysis on suspicious or anomalous files.

\subsection{Implementation and Training}
\label{subsec:plcembed_training}
We implement the convolution-attention pipeline using PyTorch. Data loading converts each PLC binary to a tensor of bytes truncated or padded to a maximum length (commonly 64K). The embedding dimension $d$ is selected based on preliminary validation experiments; a larger dimension may capture more nuanced patterns but requires more memory. Training proceeds for a fixed number of epochs (often 30), with early stopping criteria based on validation accuracy. Different classification heads can be swapped in for toolchain versus functionality tasks, but the same underlying architecture is retained.

All experiments use a train-test split at the granularity of individual ST programs to avoid overlap across data partitions. Batch sizes of 16 or 32 are typical. Hyperparameters such as learning rate and weight decay are tuned on a small validation set. Class weighting or oversampling mitigates imbalance in some of the functionality categories. During inference, the pipeline receives a raw binary, prepares the byte sequence, and outputs the predicted label.


\section{Experiments}
\label{sec:experiments}

In this section, we evaluate \textbf{PLCEmbed} on two classification tasks discussed earlier: toolchain provenance identification and functionality classification. We first describe the experimental setup, then present the results for toolchain provenance (Sec.~\ref{subsec:toolchain_results}). Results for functionality classification follow in subsequent subsections.

\subsection{Experimental Setup}
\label{subsec:exp_setup}
We use the \textit{PLC-BEAD} dataset introduced in Sec.~\ref{sec:plcbead}, containing 729 Structured Text (ST) programs compiled across four toolchains. All experiments are conducted on a server equipped with Intel Core i7-7820X CPU @3.60GHz with 16GB RAM and two NVIDIA GeForce GTX Titan Xp GPUs 

\paragraph{Evaluation Metrics.}
We report accuracy on the test set to measure how often the predicted class (toolchain or functionality) matches the ground truth. We may also include F1 scores (macro or weighted) if relevant, especially in tasks where some classes have fewer samples. In addition, we utilize the Cohen Kappa score ($\kappa$) and the Matthews correlation coefficient (MCC) to account for the disparity in class sizes and provide a fair evaluation. 


\begin{table}[!ht]
    \centering
    \caption{The Performance of \textit{PLCEmbed} on Toolchain Provenance Identification.}
    \begin{tabular}{ |p{0.7cm} | p{2.4cm} |p{0.8cm} |p{0.8cm} |p{0.8cm}| p{0.8cm} |}
        \hline
        Model & Dataset & Acc & F1 & $\kappa$ & MCC \\
        \hline
        &  \textit{PLC-BEAD} &91.77\% &91.77\% &89.01\% &89.02\% \\
        base- line& \textit{PLC-BEAD}-5\%-polluted &92.22\% &92.53\% &89.62\% &89.66\% \\
        & \textit{PLC-BEAD}-10\%-polluted &91.66\% &91.65\% &88.87\% &88.90\% \\
       
        \hline

        & \textit{PLC-BEAD} &93.11\% &93.10\% &90.80\% &90.81\% \\
        ours& \textit{PLC-BEAD}-5\%-polluted &92.66\% &92.64\% &90.21\% &90.25\% \\
        & \textit{PLC-BEAD}-10\%-polluted &92.64\% &92.64\% &90.19\% &90.21\% \\
        \hline
    \end{tabular}
    \label{tab:result_tp}
\end{table}

\subsection{Toolchain Provenance Results}
\label{subsec:toolchain_results}
We first evaluate our model on the task of identifying which compiler produced a given PLC binary. 

Table~\ref{tab:result_tp} summarizes accuracy, F1-score, and other relevant metrics on the test set. The proposed framework achieves strong performance, often above 90\% accuracy for predicting the correct compiler, with some variation across compiler types. In particular, binaries originating from OpenPLC-V3 tend to exhibit distinct byte patterns that are easier to learn, whereas CoDeSys binaries sometimes contain shared library code that can overlap with features from other toolchains. Nonetheless, the attention-based layers help capture global cues, such as the file header or compiler-specific data segments, which improves disambiguation. The CNN-only baseline performs moderately well but sometimes misclassifies binaries that share structural similarities across compilers. Without the attention-based mechanism, it may overlook long-range features that differentiate compilers more definitively.

The results indicate a substantial gain for all compilers. In practical forensic applications, even small gains in precision can significantly cut down investigative workload. A correct toolchain assignment narrows the search for known vulnerabilities and helps analysts confirm whether an executable originated from a trusted development environment. Sec.~\ref{subsec:discussion} further explores the role of this classification in ICS forensics, including how partial matches can still provide valuable clues in legacy systems with incomplete data.

\begin{table}[!ht]
    \centering
    \caption{The Performance of \textit{PLCEmbed} on Functionality Prediction.}
    \begin{tabular}{ |p{0.7cm} | p{2.5cm} |p{0.8cm}  |p{0.8cm} |p{0.8cm}| p{0.8cm} |}
        \hline
        Model & Dataset & Acc & F1 & $\kappa$ & MCC \\
        \hline
        & \textit{OpenPLC} V3 &45.18\% &41.24\%  &39.38\% &39.84\%\\
        & \textit{OpenPLC} V2 &38.77\% &36.07\%  &33.73\% &34.08\%\\

        & \textit{PLC-BEAD} &39.28\% &37.75\%  &33.70\% &33.78\%\\
        base- line & \textit{PLC-BEAD}-5\%-polluted &38.37\% &36.68\%  &32.63\% &32.75\%\\
        & \textit{PLC-BEAD}-10\%-polluted &35.65\% &33.74\%  &29.42\% &29.58\%\\
        \hline
        & \textit{OpenPLC} V3 &46.85\% &43.42\%  &41.28\% &41.73\%\\
        & \textit{OpenPLC} V2 &41.68\% &39.34\%  &37.04\% &37.36\%\\

        & \textit{PLC-BEAD} &42.28\% &40.35\%  &36.63\% &36.83\%\\
        ours& \textit{PLC-BEAD}-5\%-polluted &39.12\% &36.72\%  &33.04\% &33.29\%\\
        & \textit{PLC-BEAD}-10\%-polluted &38.23\% &35.25\%  &31.86\% &32.15\%\\
        \hline
    \end{tabular}
    \label{tab:result_func}
\end{table}

\subsection{Functionality Classification Results}
\label{subsec:func_classification}

We now evaluate both \textbf{PLCEmbed} and the \textbf{CNN-only baseline} on \textbf{functionality classification}, where each binary must be assigned to one of the 22 functionality categories from Sec.~\ref{subsubsec:labeling}. This task is typically more challenging than toolchain provenance due to potential overlaps between functionalities (for instance, math blocks that also include timing components) and uneven class distributions.

\paragraph{Overall Performance.}
Table~\ref{tab:result_func} summarizes the performance of our models on the test set. 
First, we assessed the efficacy of the models on how well they classify the functionality of binaries compiled using a single toolchain, specifically \textit{OpenPLC V3} and \textit{OpenPLC V2}, as these toolchains have the most significant number of binaries in our dataset. 
Our model achieved an accuracy of 46.85\% and 41.68\% for \textit{OpenPLC V3} and \textit{OpenPLC V2}, respectively.

While these results demonstrate the potential of our approach, it is essential to acknowledge that the performance is not exceptionally high. This could be attributed to several factors, such as the intrinsic difficulty of the task and the diversity and complexity of PLC programs. The lower performance on metrics like Cohen's Kappa score (41.28\% and 37.04\% for \textit{OpenPLC V2} and \textit{V3}, respectively) suggests that our model is affected by class imbalance in the dataset.
We also evaluated our model on the entire \textit{\textit{PLC-BEAD}-Func} dataset, which includes binaries from all four toolchains. In this setting, our model achieved an accuracy of 42.28\% and a Cohen's Kappa score of 36.63\%.

Although the baseline CNN-only model performs reasonably well for more common categories, PLCEmbed (CNN+Transformer) achieves higher accuracy overall, suggesting that the transformer’s global context helps distinguish subtler functional differences scattered across the byte sequence. Some categories with fewer samples (for example, specialized building control blocks) exhibit higher variance in both models’ predictions.

\paragraph{Imbalanced Classes.}
Certain functionalities, such as \textit{Math} or \textit{Timing}, occur more frequently than others (for example, advanced building automation). Despite label weighting and the data filtering performed in Sec.~\ref{subsubsec:labeling}, a few categories remain underrepresented. Both PLCEmbed and the baseline see lower F1 scores in these rare classes, though PLCEmbed consistently outperforms the baseline by a small margin. This indicates that local byte patterns alone may not suffice to capture complex functionalities unless supported by a mechanism (such as self-attention) that integrates information from different parts of the binary.

\paragraph{Insights.} In practical ICS scenarios, analysts can use functionality classification to verify whether an uploaded binary aligns with the intended control logic. If a malicious actor replaced a “Timing” module with a “Network” module that exfiltrates data, for instance, a high-performing classifier might flag this discrepancy even without source code. However, these results confirm that functionality classification presents an additional layer of complexity beyond toolchain detection. 
Besides class imbalance, one of the main obstacles arises when binaries compiled with the same toolchain share significant code segments, despite having different functionalities.
This can be attributed to the use of common libraries, runtime environments, and compiler-specific optimizations that result in similar binary structures across different programs.

For instance, in our dataset, the \textit{ACTUATOR\_PUMP} and \textit{TIMER\_1} programs, both compiled using \textit{OPENPLC V3}, have different functionality labels but share a notable portion of their code segments. 
The \textit{ACTUATOR\_PUMP} program implements a pump interface controllable by an input, while the \textit{TIMER\_1} program realizes a simple timer that generates output pulses on selected days. 
Conversely, the same \textit{ACTUATOR\_PUMP} program compiled using \textit{OpenPLC V3} and \textit{GEB} exhibits differences in almost all code segments, despite having similar functionality.
 
These observations highlight the need for more advanced binary analysis techniques that go beyond simple structural analysis. Future research could explore methods that consider additional features, such as control flow graphs, data flow analysis, or ML-based approaches that can learn to distinguish between different functionalities based on higher-level patterns and abstractions.

\subsection{Robustness and Additional Experiments}
\label{subsec:robustness}

We also study the model's robustness to noise. Industrial forensics sometimes involves partially corrupted binaries or incomplete memory dumps. To simulate this, we flip 5\% and 10\% of the bytes in the test-set files at random. Table~\ref{tab:result_tp} and~\ref{tab:result_func} report toolchain and functionality accuracy under these conditions. Although performance drops compared to the unmodified files, the decrease is modest (1--3\% at most), suggesting that convolution filters and attention can tolerate moderate corruption as long as key compiler or functional patterns remain intact.

\subsection{Discussion}
\label{subsec:discussion}
Our experiments indicate that \textbf{PLCEmbed} (CNN+Transformer) surpasses the simpler \textbf{CNN-only baseline} in both toolchain provenance and functionality classification tasks. This performance gap is more pronounced when classes exhibit significant overlap in byte patterns or when only limited data is available. Although pure CNNs capture local sequences effectively, the global relationships modeled by the transformer seem essential for disambiguating bytes that occur far apart in the binary.

\paragraph{Practical Implications.}
From an ICS security perspective, a data-driven method that recovers high-level properties (compiler origin or functionality) from raw binaries offers several advantages. Investigators can identify suspicious binaries that appear to originate from unexpected compilers or that do not match the stated control logic.
Moreover, maintenance teams can detect variations introduced by updated compiler versions or emerging vendor-specific features that might be invisible without proprietary documentation.

\paragraph{Limitations.}
Although these results are promising, the dataset still represents a subset of possible ICS code scenarios, and certain rare functional categories remain difficult to classify accurately. 
Potential avenues for future work include exploring more advanced model architectures, incorporating additional sources of information (e.g., control flow graphs, memory access patterns), and developing techniques to handle class imbalance and data corruption more effectively.
Nonetheless, our findings underscore the value of raw-byte, compiler-agnostic approaches for PLC binary analysis.



\section{Conclusion}
\label{sec:conclusion}
We introduced the PLC-BEAD dataset, the first comprehensive collection of PLC binaries compiled from over 700 unique programs using four diverse compilers. As our main contribution, the dataset fills a critical gap in ICS research by providing high-quality source-binary pairs and detailed functionality labels that enable systematic analysis of PLC executables. Our work demonstrates that a data-driven binary embedding framework, which combines convolutional layers with Transformer-style self-attention, can effectively leverage the rich information contained in these binaries to predict both compiler provenance and high-level functionality.

\section*{Acknowledgment}
The authors thank the colleagues of the Embedded \& Cyber-Physical Systems (AICPS) Lab, from the University of California, Irvine, particularly Qingrong Zhou, Chih-Yuan Ting, and Lelin Pan, for their contribution during the course of this research. 
This research was supported by Siemens AG. Any opinions, findings, conclusions, or recommendations expressed in this paper are those of the authors and do not necessarily reflect the views of the funding body.
\bibliographystyle{IEEEtran}
\bibliography{bibfile}

\end{document}